{
\documentclass{aa}

\usepackage{amsmath, amssymb, amsfonts, amscd}
\usepackage{mathtools} 
\usepackage{array} 
\usepackage[varg]{txfonts} 

\usepackage{graphicx} 

\usepackage{natbib}
\bibpunct{(}{)}{;}{a}{}{,} 

\begin{document}

\title{Impact of rotation on stochastic excitation\\ of gravity and gravito-inertial waves in stars}

\author{S. Mathis\inst{1,2}
\and C. Neiner\inst{2}
\and N. Tran Minh$^{\dag}$\thanks{Nguyet Tran Minh passed away on January 11,  2013.}}

\institute{
Laboratoire AIM, CEA/DSM - CNRS - Universit\'e Paris Diderot, IRFU/SAp Centre de Saclay, F-91191 Gif-sur-Yvette, France\\
\email{stephane.mathis@cea.fr}
\and LESIA, Observatoire de Paris, CNRS UMR 8109, UPMC, Univ. Paris-Diderot, 5 place Jules Janssen, 92195 Meudon, France\\
\email{coralie.neiner@obspm.fr} 
}

\date{Received ... / accepted ...}

\abstract
{Gravity waves (or their signatures) are detected in stars thanks to helio- and asteroseismology and they may play an important role in the evolution of stellar angular momentum. Moreover, the observational study of the CoRoT target HD\,51452 by Neiner and collaborators demonstrated the potential strong impact of rotation on the stochastic excitation of gravito-inertial waves in stellar interiors.}
{Our goal is to explore and unravel the action of rotation on the stochastic excitation of gravity and gravito-inertial waves in stars.}
{The dynamics of gravito-inertial waves in stellar interiors, both in radiation and in convection zones, is described with a local { non-traditional} f-plane model. Their couplings with convective turbulent flows leading to their stochastic excitation is studied in this framework.}
{First, we find that, in the super-inertial regime in which the wave frequency is over twice the rotation frequency ($\sigma>2\Omega$), the evanescence of gravito-inertial waves in convective regions decreases with decreasing wave frequency. Next, in the sub-inertial regime ($\sigma<2\Omega$), gravito-inertial waves become purely propagative inertial waves in convection zones. { Simultaneously, turbulence in convective regions is modified by rotation. Indeed, the turbulent energy cascade towards small scales is slowed down and in the case of rapid rotation, strongly anisotropic turbulent flows are obtained that can be understood as complex non-linear triadic interactions of propagative inertial waves.} These { different} behaviours, due to the action of the Coriolis acceleration, strongly modify the wave couplings with turbulent flows. { On one hand, turbulence weakly influenced by rotation is coupled with evanescent gravito-inertial waves. On the other hand, rapidly rotating turbulence is intrinsically and strongly coupled with sub-inertial waves. Finally, to study these mechanisms, the traditional approximation cannot be assumed because it does not properly treat the couplings between gravity and inertial waves in the sub-inertial regime.}}
{Our results demonstrate the action of rotation on stochastic excitation of gravity waves thanks to the Coriolis acceleration which modifies their dynamics in rapidly rotating stars { and turbulent flows}. As the ratio $2\Omega/\sigma$ increases, the couplings and thus the amplitude of stochastic gravity waves are amplified.}

\keywords{hydrodynamics -- waves -- turbulence -- stars: rotation -- stars: evolution}

\titlerunning{Impact of rotation on stochastic excitation of gravity waves}
\authorrunning{Mathis, Neiner \& Tran Minh}

\maketitle

\section{Introduction}

Gravity waves propagate in stably stratified stellar radiation zones, such as the radiative core of low-mass stars and the external radiative envelope of intermediate-mass and massive stars \citep[e.g.][]{Aertsetal2010}. When such waves (or their signatures) are detected thanks to helioseismology  \citep[][]{Garciaetal2007} and asteroseismology \citep[e.g.][]{Becketal2011,Papicsetal2012,Neineretal2012}, they constitute a powerful probe of stellar structure \citep[e.g.][]{TC2011,Beddingetal2011} and internal dynamics, for example of differential rotation \citep[][]{Garciaetal2007,Becketal2012,Deheuvelsetal2012,Mosseretal2012}. Furthermore, when they propagate, gravity waves are able to transport and deposit a net amount of angular momentum because of their radiative damping and corotation resonances \cite[e.g.][]{GoldreichNicholson1989,Schatzman1993,ZahnTalonMatias1997}. Therefore, they are invoked, together with magnetic torques, to explain the quasi-uniform rotation until 0.2R$_{\odot}$ in the solar radiative core \citep[][]{CharbonnelTalon2005}, the mixing of light elements in low-mass stars \citep[][]{TalonCharbonnel2005}, the weak differential rotation in subgiant and red giant stars \citep[][]{Cellieretal2012}, and the transport of angular momentum necessary to explain mass-loss in active massive stars such as Be stars \citep[][]{Huatetal2009,Neineretal2013,LeeSaio1993,Lee2013}. Thus, it is necessary to get a good understanding of their excitation mechanisms and a precise prediction of their amplitude.

In single stars, two mechanisms can excite gravity waves: the $\kappa$-mechanism due to opacity bumps \citep[e.g.][]{Unnoetal1989,GastineDintrans2008} and stochastic motions both in the bulk of convective regions and at their interfaces with adjacent radiation zones where turbulent convective structures (plumes) penetrate because of their inertia \citep[e.g.][]{Press1981,Browningetal2004,RogersGlatzmaier2005,Belkacemetal2009,Cantielloetal2009,Samadietal2010,BMT2011,Rogersetal2012,Shiodeetal2013}. In this work, we focus on stochastic excitation.\\

In addition to stochastically excited mixed gravito-acoustic modes currently detected with space asteroseismology, e.g. in red giants \citep[e.g.][]{Becketal2011}, a detection has been obtained thanks to CoRoT in the hot Be star HD\,51452 \citep{Neineretal2012}. In this star, which rotates close to its critical angular velocity, authors discovered gravity modes strongly influenced by the Coriolis acceleration, i.e. gravito-inertial modes \citep[][]{DintransRieutord2000,Mathis2009,Ballotetal2010}. They propose that these modes are probably excited stochastically by turbulent convection in the core or/and in the subsurface convection zone, since this star is too hot to excite gravity modes with the $\kappa$-mechanism. Moreover, the detected gravito-inertial modes with the largest amplitudes have frequencies below the inertial one at $2\Omega$ (where $\Omega$ is the angular velocity of the star), which are the most influenced by the Coriolis acceleration. Therefore, this discovery points the potential important action of rotation on stochatic excitation of gravity and gravito-inertial waves (hereafter GIWs) in stellar interiors. This has been poorly explored until now \citep[][]{BelkacemMathisetal2009,Rogersetal2013}.

To identify the associated mechanisms and unravel the related signatures, we choose here to generalise the work by \cite{GoldreichKumar1990} and \cite{LecoanetQuataert2013} taking the action of the Coriolis acceleration into account. First, in Sect.~\ref{giw}, we present the local rotating set-up in which we describe the different regimes of the dynamics of GIWs both in radiation and in convection zones. Next, in Sect.~\ref{turb}, their stochastic excitation by turbulent convective flows is studied { with a particular focus on the effects of slow and rapid rotation}. Finally, in Sect.~\ref{discuss}, we conclude on the impact of rotation on the stochastic excitation of GIWs in stars and we discuss the consequences for asteroseismology and for the study of their angular momentum evolution.  
 
\section{Gravito-inertial waves in stellar interiors}\label{giw}

The first step to explore the impact of rotation on GIWs stochastic excitation in stellar interiors is to describe properly their different propagation regimes as a function of their frequencies. 

\subsection{The rotating ``f-plane'' reference frame}

\begin{figure}[t!]
\centering
\includegraphics[width=0.425\textwidth]{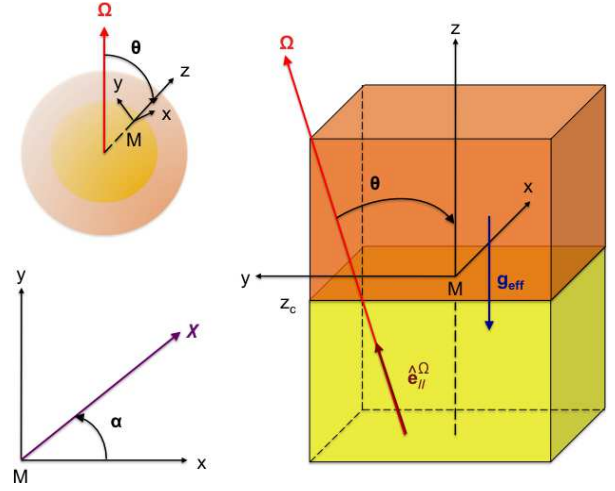}        
\caption{Local studied "$f$-plane" reference frame { (${\vec{\hat e}}_{/\!/}^{\Omega}$ is the unit-vector along the rotation axis)}. Here, the radiative and convective regions are respectively in yellow and orange. This configuration corresponds to the case of low-mass stars{ ; $z_c$ is the altitude of the transition radiation/convection}. The vertical structure of this set-up can be inverted to treat the case of intermediate-mass and massive stars.}
\label{Fig1MN}
\end{figure}

To reach our objective, we follow the local approach adopted by \cite{GoldreichKumar1990} and \cite{LecoanetQuataert2013} but taking rotation into account. We thus choose to consider a cartesian region, centered on a point M of a radiation-convection interface, where $\Theta$ is the angle between the local effective gravity ${\vec g}_{\rm eff}$\footnote{This effective gravity is the sum of the self-gravity ${\vec g}$ and of the centrifugal acceleration $1/2\,\Omega^2{\vec\nabla}S^2$, where $S$ is the distance from the rotation axis.} and the rotation vector $\vec\Omega$ (see Fig.~\ref{Fig1MN}). M$x$, M$y$ and M$z$ are the axes along the local azimuthal, latitudinal and vertical (along ${\vec g}_{\rm eff}$) directions respectively. { We define $z_c$, the altitude of the transition between the radiation and convection zones}. This so-called ``$f$-plane'' \citep{Pedlosky1982} is co-rotating with the stellar angular velocity $\vec\Omega$. To write equations of wave dynamics in this reference frame, we then introduce the two components of this vector
\begin{equation}
f=2\Omega\cos\Theta\hbox{ }\hbox{ }\hbox{and}\hbox{ }\hbox{ }{\widetilde f}=2\Omega\sin\Theta,
\end{equation} 
along the vertical and latitudinal directions, respectively. Following \cite{GerkemaS2005}, both components are taken into account to ensure a complete and correct treatment of GIWs' dynamics, both in the radiation and in the convective regions. Assuming constant $f$ gives the so-called ``non-traditional $f$-plane'', contrary to the usual traditional approximation in which the horizontal component $\widetilde f$ is neglected \citep[][]{Eckart1961}. Taking ${\widetilde f}$ into account allows to correctly treat the coupling between GIWs and inertial waves \citep{Gerkemaetal2008}.

\subsection{The Poincar\'e equation}

To get the equation that governs GIWs' dynamics, the Poincar\'e equation, we write the linearised equations of motion of the stellar stratified fluid on the non-traditional $f$-plane assuming the Boussinesq and the Cowling approximations \citep{GerkemaS2005,Cowling1941}. First, we introduce the GIWs' velocity field $\vec u=\left(u,v,w\right)$, where $u$, $v$ and $w$ are the components in the local azimuthal, latitudinal and vertical directions respectively. Next, we define the fluid buoyancy
\begin{equation}
b=-{\overline g_{\rm eff}\left(z\right)}\frac{\rho'\left(\vec r,t\right)}{{{\overline \rho}\left(z\right)}}
\end{equation}
and the corresponding Brunt-V\"ais\"al\"a frequency
\begin{equation}
N^2\left(z\right)=-\frac{{\overline g_{\rm eff}}}{{\overline \rho}}\,\frac{{\rm d}{\overline\rho}}{{\rm d}z},
\end{equation}
where $\rho'$ and $\overline\rho$ are the density fluctuation and the reference background density, and $t$ is the time. The three linearised components of the momentum equation are given by:
\begin{equation}
\left\lbrace
\begin{array}{lcl}
\partial_{t}w-{\widetilde f} u=-\partial_{z}p+b\\
\partial_{t}v+f u=-\partial_{y}p\\
\partial_{t}u-f v+{\widetilde f}w=-\partial_{x}p
\end{array}\right.\,.
\end{equation}
Next, we write the continuity equation in the Boussinesq approximation
\begin{equation}
\partial_{z}w+\partial_{y}v+\partial_{x}u=0.
\end{equation}
Finally, we get the equation for energy conservation in the adiabatic limit
\begin{equation}
\partial_{t}b+N^{2}\left(z\right) w=0.
\end{equation}
Eliminating the horizontal components of the velocity, the pressure and the buoyancy, we reduce the system to an equation for the vertical velocity
\begin{equation}
\partial_{t,t}\left[{\nabla}^{2} w\right]+4\left({\vec\Omega}\cdot{\vec\nabla}\right)^{2}w+N^2{\nabla}_{H}^{2}w=0,
\label{M1}
\end{equation}
where ${\nabla}_{H}^{2}$ is the horizontal Laplacian. We then consider a given monochromatic GIW with a frequency $\sigma$ that propagates in the direction $\left(\cos\alpha,\sin\alpha\right)$ in the (M$xy$) plane (see Fig. \ref{Fig1MN}). Introducing the reduced horizontal coordinate $\chi=x\cos\alpha+y\sin\alpha$ (with $\partial_{x}=\cos\alpha\,\partial_{\chi}$ and $\partial_{y}=\sin\alpha\,\partial_{\chi}$) and substituting $w\left(\vec r,t\right)=W\left(\vec r\right)\exp\left[i\sigma\,t\right]$, we finally get the Poincar\'e equation for GIWs:
\begin{equation}
A\left(z\right)\,\partial_{\chi,\chi}W+2B\,\partial_{\chi,z}W+C\,\partial_{z,z}W=0,
\label{M2}
\end{equation}
where 
\begin{equation}
\left\lbrace
\begin{array}{lcl}
A=N^{2}\left(z\right)-\sigma^2+f_{s}^{2}\hbox{ }\hbox{ }\hbox{with}\hbox{ }\hbox{ }f_{s}={\widetilde f}\sin\alpha\\
B=f f_{s}\\
C=f^2-\sigma^2
\end{array}\right.\,.
\end{equation}

Moreover, even if GIWs' dynamics constitutes a bidimensionnal problem because of the mixed derivative $\partial_{\chi,z}$ \citep[e.g.][]{DintransRieutord2000}, it is possible in our set-up to introduce the transformation  
\begin{equation}
W=\Psi\left(z\right)\exp\left[i k_{\perp}\left(\chi+{\tilde\delta} z\right)\right]\hbox{ }\hbox{ }\hbox{where}\hbox{ }\hbox{ }{\tilde\delta}=-\frac{B}{C},
\label{vert}
\end{equation}
which leads to
\begin{equation}
\frac{{\rm d}^2}{{\rm d}z^2}\Psi+k_{V}^{2}\left(z\right)\Psi=0,
\label{Schro}
\end{equation}
where
\begin{equation}
k_{V}^{2}\left(z\right)=k_{\perp}^2\left[\frac{B^2-AC}{C^2}\right]=k_{\perp}^2\left[\frac{N^2\left(z\right)-\sigma^2}{\sigma^2-f^2}+\left(\frac{\sigma\,f_s}{\sigma^2-f^2}\right)^2\right],
\label{M3}
\end{equation}
$k_{\perp}$ being the wave number in the $\chi$ direction \citep{GerkemaS2005}. This enables us to use the method of vertical modes as in the non-rotating case and the associated tools since, as demonstrated in \cite{GerkemaS2005}, the modal functions $W_{j}$ that verify boundary conditions constitute an orthogonal and complete basis. 

\subsection{Gravito-inertial wave propagation in rotating stars}

We will now describe GIWs propagation as a function of their frequency $\sigma$. 

At a given $z$, GIWs are propagative if $\Delta=B^2-AC>0$. This leads to an allowed frequency spectrum $\sigma_{-}<\sigma<\sigma_{+}$, where 
\begin{equation}
\sigma_{\pm}=\frac{1}{\sqrt 2}\sqrt{\left[N^2+f^2+f_{s}^{2}\right]\pm\sqrt{\left[N^2+f^2+f_{s}^{2}\right]^{2}-\left(2fN\right)^{2}}}.
\end{equation}
In convective regions, which excite GIWs, the local vertical number (Eq. (\ref{M3})) becomes
\begin{equation}
k^{2}_{\rm CZ}\equiv k_{\perp}^2\frac{\sigma^2}{\left(\sigma^2-f^2\right)^2}\left[\left(f^2+f_{s}^{2}\right)-\sigma^2\right]=k_{\perp}^{2}\frac{\left[\left[{{\widetilde R}_{\rm o}}\left(\Theta,\alpha\right)\right]^{-2}-1\right]}{\left(1-R_{\rm o}^{-2}\cos^2\Theta\right)^{2}},
\label{M4}
\end{equation}
where we defined a local wave Rossby number
\begin{equation}
{\widetilde R}_{\rm o}=R_{\rm o}\left[\cos^{2}\Theta+\sin^{2}\alpha\sin^{2}\Theta\right]^{-1/2}
\end{equation}
expressed as a function of the wave's Rossby number $R_{\rm o}=\sigma/2\Omega$.

Then, we identify for $\alpha=\pi/2$ the allowed frequency domain for inertial waves $0<\sigma<2\Omega$ (i.e. { $R_{\rm o}<1$}). On one hand, in the local super-inertial regime ({ ${\widetilde R}_{\rm o}>1$}), there are turning points ($z_{t;i}$) in the radiation zones for which $k_{V}^{2}\left(z_{t;i}\right)=0$ (see Eq. (\ref{TP})) and GIWs are evanescent in convection zones. On the other hand, in the local sub-inertial domain ({ ${\widetilde R}_{\rm o}<1$}), GIWs propagate in the whole radiation zone and become propagative inertial waves in convection zones. 

\begin{figure}[t!]
\centering
\includegraphics[width=0.495\textwidth]{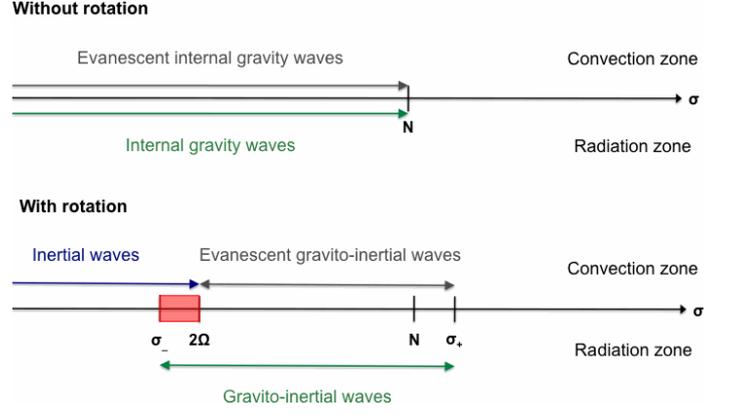}        
\caption{Low frequency spectrum of waves in a non-rotating ($\Omega=0$, top) and in a rotating (bottom) star. For each case, waves in the convection and radiation zones are indicated at the top and bottom, respectively. { The red box corresponds to sub-inertial gravito-inertial waves where the cavities of gravity and inertial waves are coupled.}}
\label{Fig2MN}
\end{figure}

\begin{figure*}[t!]
\centering  
\includegraphics[width=\textwidth]{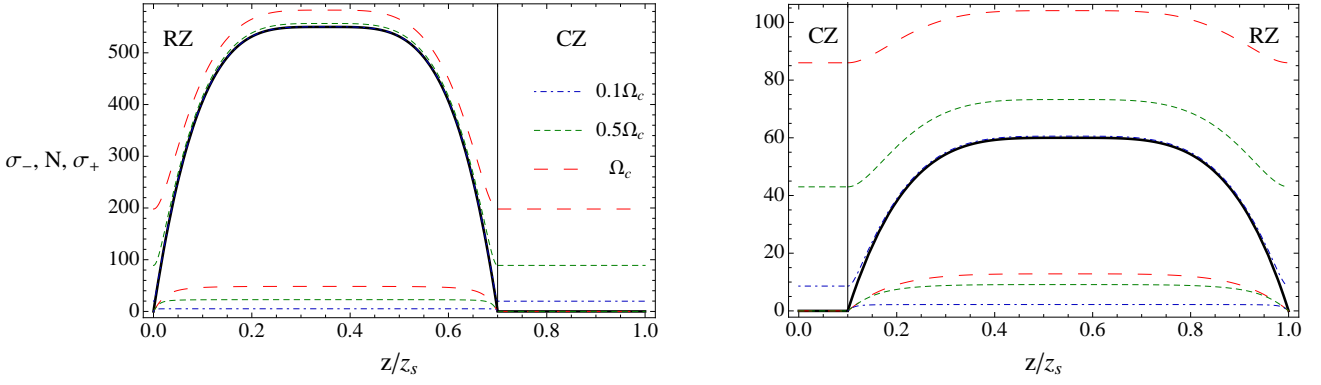}      
\caption{Synthetic profiles of $\left\{\sigma_{-},\sigma_{+}\right\}$ (color lines) and $N$ (solid black line) for a $1M_{\odot}$ solar-type star (left) and a $12M_{\odot}$ massive star (right) for $\alpha=\pi/2$, $\Theta=5\pi/12$, and various rotation rates: $\Omega=0.1\Omega_{c}$ (blue { dotted-dashed} line), $0.5\Omega_{c}$ ({ green} dashed line), $1\Omega_{c}$ (red long-dashed line). The border between the convection and radiation zones (CZ and RZ) is given by the grey vertical line. In the case of the $12M_{\odot}$ massive star, note that we have filtered out the bumps of $N$ at the top of the convective core and just below the stellar surface. { In convection zones, $\sigma_{-}\sim N\sim 0$.}}
\label{Fig3MN}
\end{figure*}

These two behaviours (see Fig. \ref{Fig2MN}) have been isolated by \cite{DintransRieutord2000} for a 1.5$M_{\odot}$ star (see their Fig. 6). The cases of typical solar-type stars (1$M_{\odot}$) and massive stars (12$M_{\odot}$) are illustrated in Fig.~\ref{Fig3MN} for a small horizontal cross-section box along a given radius. By defining a synthetic profile of positive Brunt-Va\"is\"al\"a frequency with a maximum value $N_{\rm max}$ computed using a 1-D stellar evolution code, we plot $\left\{\sigma_{-},\sigma_{+}\right\}$ profiles for various values of the rotation rate: $\Omega=0.1\Omega_{\rm c}$ (in blue), $0.5\Omega_{\rm c}$ (in purple) and $\Omega_{\rm c}$ (in red). The critical angular velocity is  $\Omega_{\rm c}=\sqrt{GM/R^3}$, where $G$, $M$ and $R$ are the gravity universal constant, the stellar mass and radius, respectively. In the case of the $12M_{\odot}$ massive star, note that we have filtered out the bumps of $N$ at the top of the convective core and just below the stellar surface. We point out that $S_{\Omega}=N_{\rm max}/{2\Omega}$ $\left(\hbox{and}\hbox{ }S_{\Omega;c}=N_{\rm max}/2\Omega_{c}\right)$ are the control parameters that determine if the second regime, in which GIWs propagate in the whole radiation zone and become inertial waves in the convection zone, spreads over a large frequency range. As shown in Fig.~\ref{Fig3MN}, this can be the case of rapidly rotating massive stars. Therefore, we have from now on to distinguish these two regimes.

\subsubsection{Super-inertial regime $\left({\widetilde R}_{\rm o}>1\right)$}

In the case of the super-inertial regime, there are two turning points ($z_{t;1}$, $z_{t;2}$ with $z_{t;1}<z_{t;2}$) in the radiation zone for which $k_{V}^{2}=0$ that corresponds to values of $z$ where
\begin{equation}
N^2=\sigma^2\frac{\left[\sigma^2-\left(f^2+f_{s}^{2}\right)\right]}{\left(\sigma^2-f^2\right)}.
\label{TP}
\end{equation}

To get an asymptotic expression of $\Psi$ (defined in Eq.~(\ref{vert})) in the whole radiation zone, we have to go beyond the usual JWKB approximation \citep{FF2005}, which fails at turning points, and adopt the uniform Airy approximation \citep{BM1972}. The latter has been introduced in quantum mechanics and allows us to get asymptotic expressions in terms of Airy functions, which are valid close and far from turning points. Following \cite{BVT1988}, we obtain in the case of an external convective zone:
\begin{equation}
\Psi\left(z<z_{\rm c}\right)=A\left(\frac{\Phi\left(z\right)}{k_{V}^{2}\left(r\right)}\right)^{1/4}{\rm Ai}\left[\Phi\left(z\right)\right],
\end{equation}
where
\begin{equation}
\Phi\left(z\right)=\varepsilon\left(\frac{3}{2}\int_{z_{t;2}}^{z}\vert k_{V}\left(z'\right)\vert{\rm d}z'\right)^{2/3}\hbox{ }\hbox{with}\hbox{ }\varepsilon=
\begin{cases}
1\hbox{ }\hbox{ }\hbox{if}\hbox{ }\hbox{ }k_{V}^{2}<0\\
-1\hbox{ }\hbox{ }\hbox{if}\hbox{ }\hbox{ }k_{V}^{2}>0
\end{cases}
\end{equation}
and $k_V$ given in Eq. (\ref{M3}). In the convective region, GIWs are evanescent and we obtain using asymptotic properties of Airy functions and Eq.(\ref{M4}):
\begin{equation}
\Psi\left(z>z_{\rm c}\right)=\frac{A}{\sqrt{k_{\perp}{p}}}\exp\left[-k_{\perp}{p}\left(z-z_{\rm c}\right)-\Delta\left(z_{t;2},z_{c}\right)\right],
\end{equation}
where we define
\begin{equation}
{p}={\frac{\sqrt{\vert 1-{\widetilde R}_{\rm o}^{-2}\vert}}{\vert1-R_{\rm o}^{-2}\cos^{2}\Theta\vert}}\hbox{ }\hbox{ }\hbox{and}\hbox{ }\hbox{ }\Delta\left(z_{1},z_{2}\right)=\int_{z_1}^{z_2}\vert k_{V}\vert{\rm d}z'.
\label{DampDelta}
\end{equation}
{ We recall that $z_c$ is the altitude of the transition radiation/convection.} In this regime, $p$ corresponds to the decay rate of the velocity in the convection zone. If we choose to verify no-slip boundary conditions $w\left(z_b,\chi,t\right)=w\left(z_s,\chi,t\right)=0$, { where $z_b$ and $z_s$ are the altitudes corresponding to the bottom and the top of the box}, we can deduce eigenfrequencies using Bohr quantisation \citep[][]{LandauLifshitz1965}
$
\displaystyle{\int_{z_{t;1}}^{z_{t;2}}\vert k_{V}\vert\,{\rm d}z'=\left(k+\frac{1}{2}\right)\pi}
$, 
where $k$ is an integer. 

\begin{figure}[t!]
\centering  
\includegraphics[width=0.495\textwidth]{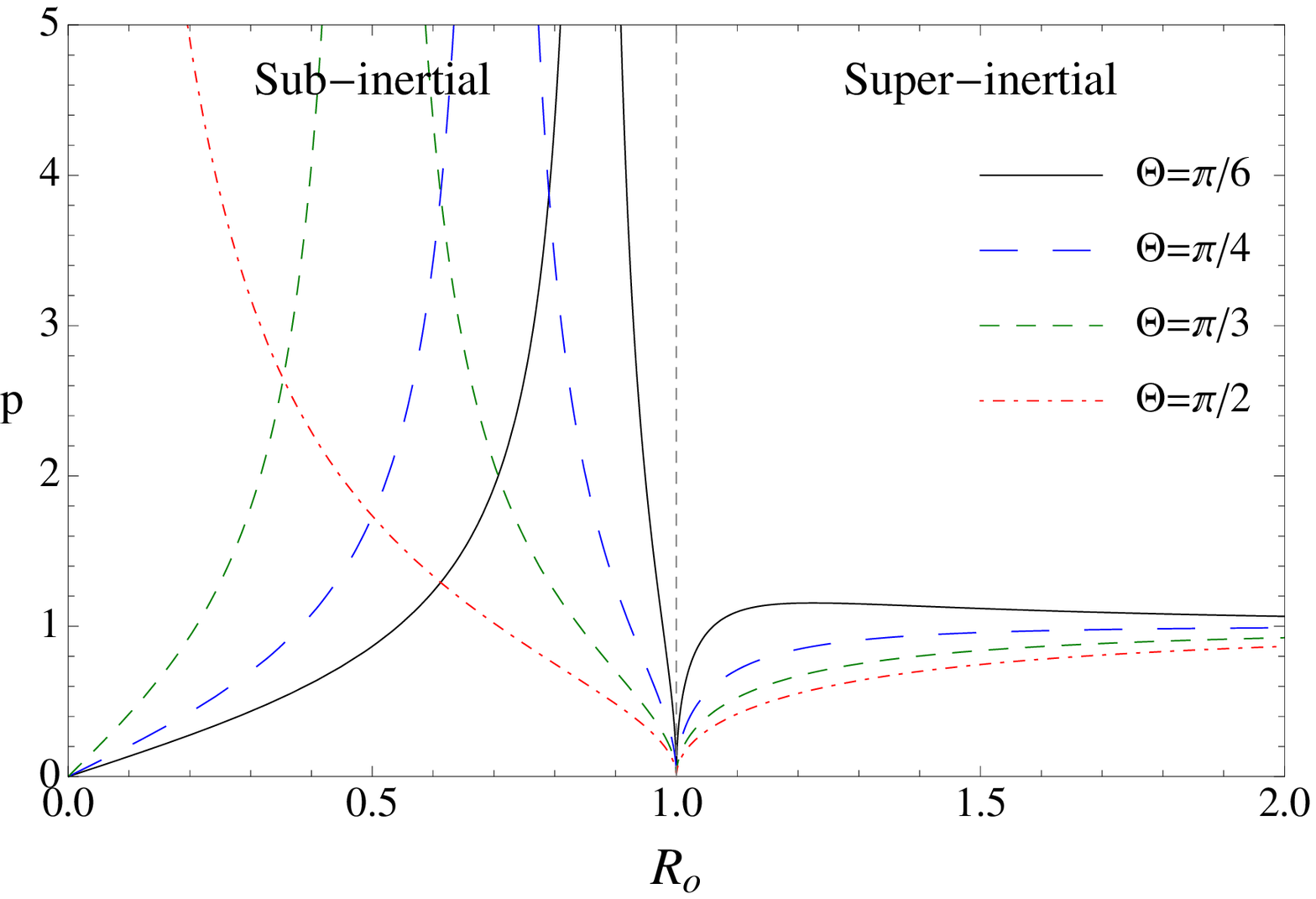} 
\includegraphics[width=0.495\textwidth]{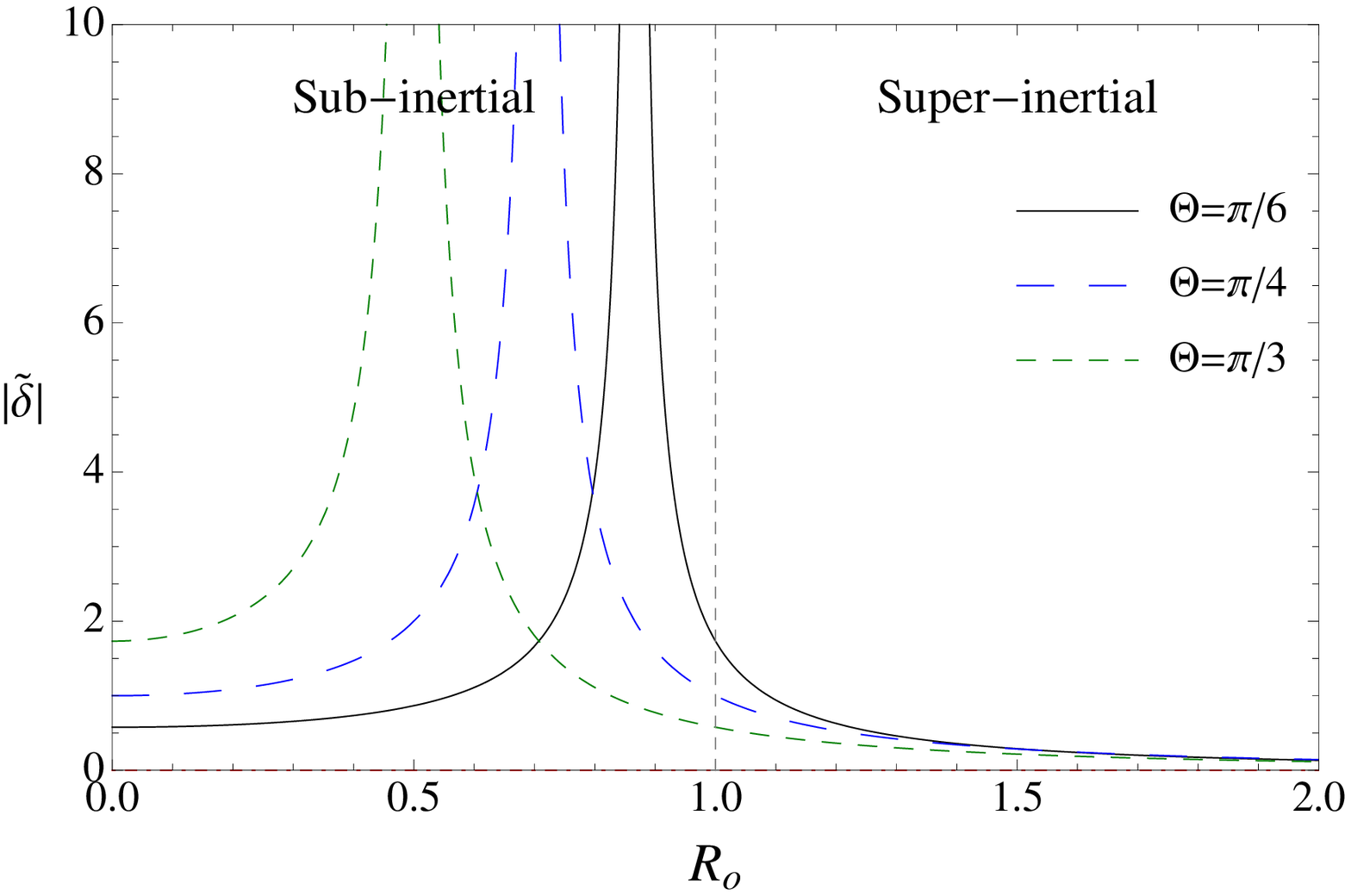}  
\caption{Evolution of $p$ (top) and ${\tilde\delta}$ (bottom) as a function of $R_{\rm o}$ for $\alpha=\pi/2$ and different inclination angle: $\Theta=\pi/6$ (black solid line), $\pi/4$ (blue long-dashed line), $\pi/3$ ({ green} dashed line) and $\pi/2$ (red dot-dashed line). ${\tilde\delta}$ vanishes for $\Theta=\pi/2$ (see Eq. (\ref{tildedelta})). The border between the sub-inertial and the super-inertial regimes is given by the grey dashed line.}
\label{Fig4MN}
\end{figure}

Then, the complete bidimensionnal expression for the vertical velocity is obtained using Eq. (\ref{vert}):
\begin{eqnarray}
W\left(z>z_{\rm c},\chi\right)&=&\frac{A}{\sqrt{k_{\perp}{p}}}\exp\left[-k_{\perp}{p}\left(R_{\rm o}\right)\left(z-z_{\rm c}\right)-\Delta\left(z_{t;2},z_{c}\right)\right]\nonumber\\
&&\times\cos\left[k_{\perp}\left(\chi+{\tilde\delta}\left(R_{\rm o}\right)z\right)\right],
\label{bidisuper}
\end{eqnarray}
where
\begin{equation}
{\tilde\delta}=-\frac{R_{\rm o}^{-2}\cos\Theta\sin\Theta\sin\alpha}{R_{\rm o}^{-2}\cos^{2}\Theta-1}.
\label{tildedelta}
\end{equation}
The evolution of $p$ and ${\tilde\delta}$ as a function of $R_{\rm o}$ and $\Theta$ are plotted in Fig. \ref{Fig4MN}. In the super-inertial regime, $p$ increases with $R_{\rm o}$ (with $p\rightarrow1$ for $R_{\rm o}\rightarrow\infty$). This means that the decay rate of the wave function in the convective zone decreases when the rotation rate increases until $R_{\rm o}=1$. Moreover, ${\tilde\delta}$ tends to vanish at large $R_{\rm o}$ while it increases with the rotation rate until $R_{\rm o}=1$. This corresponds to the fact that, as soon as rotation becomes important, the problem is non-separable in $z$ and $\chi$ while it is completely separable in the non-rotating case (i.e. $R_{\rm o}\rightarrow\infty$). Accordingly, modifications of GIW velocity fields occur. This shows the important impact of the Coriolis acceleration, which will modify their couplings with convective flows.

{ In this work, we use the uniform Airy approximation. In the case of very steep transition from convection to radiation on a characteristic distance $d$ such that $k_{\perp}d<1$, \cite{LecoanetQuataert2013} pointed that it would be better to adopt solutions of Eq. (\ref{Schro}) corresponding to a buoyancy frequency profile expressed as a function of an hyperbolic tangent. Such combination of steep transition and rotation will be examined in a near future and is beyond the scope of this work. In fact, the evanescent or propagative behaviors of GIWs in convection zones depends only on the sign of $k_{V}^{2}$.}

\subsubsection{Sub-inertial regime $\left({\widetilde R}_{\rm o}<1\right)$}

In the case of the sub-inertial regime, waves are propagative in the whole domain without any turning point ($k_{V}^{2}>0$). Assuming the JWKB approximation in the radiation zone, we can write
\begin{equation}
\Psi\left(z\right)=A\frac{1}{\sqrt{\vert k_{V}\vert}}\sin\left(\int_{z_b}^{z} \vert k_{V}\vert\,{\rm d}z'\right),
\end{equation}
where $k_V$ is given in Eq. (\ref{M3}). In the convective region, we get for inertial waves using Eq. (\ref{M4}):
\begin{equation}
\Psi\left(z>z_{\rm c}\right)=\frac{A}{\sqrt{k_{\perp}p}}\sin\left[k_{\perp}p\left(z-z_{\rm c}\right)+\Delta\left(z_b,z_c\right)\right],
\end{equation}
where $p$ and $\Delta$ have been defined in Eq. (\ref{DampDelta}). As in the super-inertial case, we can use Bohr quantisation to get eigenfrequencies in the case of no-slip boundary conditions, i.e. 
$
\displaystyle{\int_{z_b}^{z_s}\vert k_{V}\vert\,{\rm d}z'=n\pi}
$, where $n$ is an integer.

\begin{figure}[t!]
\centering  
\includegraphics[width=0.495\textwidth]{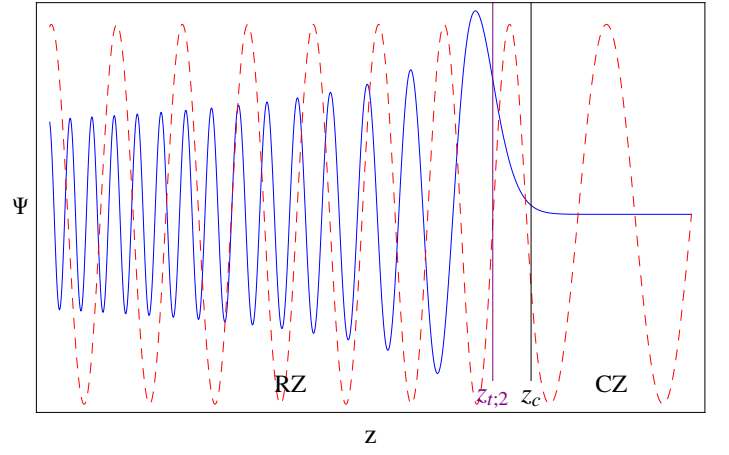}  
\caption{{ Cartoon for the general} behaviours of $\Psi$ in the super-inertial (solid blue line) and in the sub-inertial (red dashed line) regimes around the external turning point $z_{t;2}$ (vertical purple line) for a star with a convective envelope (for $z>z_{\rm c}$). The border between the convection and radiation zones (CZ and RZ) is given by the black vertical line.}
\label{Fig5MN}
\end{figure}

The complete bidimensionnal expression for the vertical velocity is obtained using Eq. (\ref{vert}):
\begin{eqnarray}
W\left(z>z_{\rm c},\chi\right)&=&\frac{A}{\sqrt{k_{\perp}{p}}}\sin\left[k_{\perp}p\left(R_{\rm o}\right)\left(z-z_{\rm c}\right)+\Delta\left(z_b,z_c\right)\right]\nonumber\\
&&\times\cos\left[k_{\perp}\left(\chi+{\tilde\delta}\left(R_{\rm o}\right)z\right)\right].
\label{bidisub}
\end{eqnarray}
In this regime, $p$ (see Eq. (\ref{DampDelta})) is the vertical wave number of the propagative inertial wave in the convection zone. The variations of $p$ and ${\tilde\delta}$ as a function of $R_{\rm o}$ in the sub-inertial regime, shown in Fig. \ref{Fig4MN}, imply that the velocity field becomes more and more oscillatory as the rotation rate increases until $R_{\rm o}=\cos(\Theta)$ where the denominators of $p$ and ${\tilde\delta}$ vanish. $\Theta_{\rm c}={\cos}^{-1}\left(R_{\rm o}\right)$ corresponds to the critical colatitude above which GIWs are trapped in the radiation zone in the sub-inertial regime \citep[e.g.][]{MdB2012}. This behaviour is strongly different from the one obtained in the super-inertial regime and will modify GIWs stochastic excitation by convective flows, as represented in Fig \ref{Fig5MN}.

\section{Gravito-inertial wave stochastic excitation by turbulent convection}\label{turb}

\subsection{ The forced Poincar\'e equation}

To obtain the amplitude of GIWs excited by turbulent convective zones, we follow \cite{GoldreichKumar1990} and \cite{LecoanetQuataert2013}. The equation for the vertical component of the wave velocity ($w$) is given for the radiation zone in  Eq. (\ref{M1}) and for the convective region by
\begin{equation}
\partial_{t,t}\left[\nabla^{2}w\right]+4\left({\vec\Omega}\cdot{\vec\nabla}\right)^{2}w=\partial_{t}{\mathcal S}=\partial_{t}\left[\partial_{z}{\vec\nabla}\cdot{\vec F}-\nabla^{2}F_{z}\right].
\label{InH}
\end{equation}
We introduce the turbulent source function $\mathcal S$, defined as a function of the convective Reynolds stresses ${\vec F}={\vec\nabla}\cdot\left({\vec u}_{\rm c}{\vec u}_{\rm c}\right)$. A complete overview of the different excitation sources, which can excite GIWs in a rotating star, has been given in \citeauthor{BelkacemMathisetal2009} \citep[2009a, see also][]{SamadiGoupil2001}. They demonstrated that we can focus on this term $\partial_{t}\mathcal S$. 

Using the orthogonality of GIW functions in our cartesian set-up, which has been demonstrated by \cite{GerkemaS2005}, we can derive a formal expression for the amplitude of GIWs depending on their super- or sub-inertial behaviour. Following \cite{GoldreichKumar1990} and \cite{LecoanetQuataert2013}{ , we expand the solution of Eq. (\ref{InH}) as
\begin{equation}
W=\frac{1}{\sqrt{\mathcal A}}\sum_{\sigma_i}A\left(t,\sigma_i\right)W_{i}\exp\left(i\sigma_i t\right),
\end{equation}
where $\sigma_i$ are the eigenfrequencies, $W_{i}$ the corresponding eigenfunctions of the homogeneous equation and $\mathcal A$ the horizontal cross-section of the box. We obtain
\begin{eqnarray}
\left| A_{\hbox{sup}}\right|&=&\frac{\sqrt {\mathcal A}\left|\sigma^2-f^2\right|}{2 \sigma\ k_{\perp}^{2}}\frac{1}{\left|\mathcal C\right|}\nonumber\\
&\times&\frac{1}{\sqrt{k_{\perp}{p}}}\int_{-\infty}^{\,t}{\rm d}\tau \int_{\mathcal V}{\rm d}x\,{\rm d}y\,{\rm d}{\zeta}\,\partial_{t}{\mathcal S}\left(x,y,\zeta,\tau\right)\nonumber\\
&\times&\exp\left[-i\,k_{\perp}\left[\left(x\cos\alpha+y\sin\alpha\right)+{\tilde\delta}\left(R_{\rm o}\right)\zeta\right]-i\sigma\tau\right]\nonumber\\
&\times&\exp\left[-k_{\perp}{p\left(R_{\rm o}\right)}\left(\zeta-z_{\rm c}\right)-\Delta\left(z_{t;2},z_{c}\right)\right]
\label{FA1}
\end{eqnarray}
and
\begin{eqnarray}
\left| A_{\hbox{sub}}\right|&=&\frac{\sqrt {\mathcal A}\left|\sigma^2-f^2\right|}{2 \sigma\ k_{\perp}^{2}}\frac{1}{\left|\mathcal C\right|}\nonumber\\
&\times&\frac{1}{\sqrt{k_{\perp}{p}}}\int_{-\infty}^{\,t}{\rm d}\tau \int_{\mathcal V}{\rm d}x\,{\rm d}y\,{\rm d}{\zeta}\,\partial_{t}{\mathcal S}\left(x,y,\zeta,\tau\right)\nonumber\\
&\times&\exp\left[-i\,k_{\perp}\left[\left(x\cos\alpha+y\sin\alpha\right)+{\tilde\delta}\left(R_{\rm o}\right)\zeta\right]-i\sigma\tau\right]\nonumber\\
&\times&\sin\left[k_{\perp}p\left(R_{\rm o}\right)\left(\zeta-z_{\rm c}\right)+\Delta\left(z_b,z_c\right)\right],
\label{FA2}
\end{eqnarray}
where $\mathcal V$ is the volume of the box. The normalisation coefficient \citep{GerkemaS2005} 
\begin{equation}
{\mathcal C}\equiv\left[\int{\rm d}\chi\int_{z_b}^{z_s}\left[N^{2}\left(z\right)-R^2\left(z\right)\right]W_{i}\left(z,\chi\right)W_{i}^{*}\left(z,\chi\right){\rm d}z\right],
\end{equation}
where
\begin{equation}
R\left(z\right)^{2}=\frac{\sigma^2f_{s}^{2}+\sigma^{2}\left(f^2-\sigma^2\right)}{f^2-\sigma^2}
\end{equation}
is introduced because the eigenfunctions $W_{i}$ are only orthogonal. Applying Eqs. (\ref{FA1}-\ref{FA2}) to the non-rotating case and working with the vertical displacement $\xi_{z}$, where $w=\partial_{t}\xi_{z}$, we recover Eq. (C3) in \cite{LecoanetQuataert2013}.} This formal expressions demonstrate how the coupling between the wave's velocity and convective turbulence is a function of $R_{\rm o}=\sigma/2\Omega$ (via $p$, see Eq.~(\ref{DampDelta}), and $\delta$, see Eq.~(\ref{tildedelta})). In the super-inertial regime, turbulent convective flows are correlated with an evanescent gravito-inertial wave (the last exponential term), while in the sub-inertial regime they couple with a propagative inertial wave (the sinusoidal term). To go further, we have to be more specific on the turbulent source function, { and to discuss its modification by rotation}.

\subsection{ The modification of turbulence by rotation}

\subsubsection{ The various regimes}

{ Rotation strongly modifies the properties of turbulent flows. First, it induces anisotropy of the flows with a tendency to a bi-dimensionalisation. Next, the turbulent energy cascade towards small scales is slowed down, in particular in the case of decaying turbulence. Finally, relatively strong columnar structures along the direction of the axis of rotation appear.

We consider the non-linear Navier-Stockes equation in the rotating reference frame for an homogeneous fluid:
\begin{equation}
\partial_{t}{\vec u}+\left({\vec u}\cdot{\vec\nabla}\right){\vec u}+2{\vec\Omega}\times{\vec u}=-\frac{1}{\rho}{\vec \nabla}P+{\vec b}+\nu{\vec\nabla}^{2}{\vec u},
\label{Eq_NL}
\end{equation}
where ${\vec b}$ corresponds to the buoyancy. If we introduce the characteristic scales of velocity ($U$), length ($L$) and time ($T$) to { non-dimensionalised} Eq. (\ref{Eq_NL}), we obtain
\begin{equation}
R_{\rm o}\partial_{t^{*}}{\vec u^{*}}+R_{\rm o}^{\rm NL}\left({\vec u^{*}\cdot{\vec\nabla}^{*}}\right){\vec u^{*}}+{\vec {\hat e}}^{\Omega}_{/\!/}\times{\vec u^{*}}= -{\vec\nabla}\pi^{*} +{\vec b}^{*}+\frac{R_{\rm o}^{\rm NL}}{R_{\rm e}}{\vec\nabla^{*}}^{2}{\vec u^{*}},
\end{equation}
where $t^{*}=t/T$, ${\vec\nabla}^{*}=L{\vec\nabla}$, ${\vec u^{*}}={\vec u}/U$, $\pi^{*}=P/\left(\rho 2\Omega L U\right)$, ${\vec b}^{*}={\vec b}/\left(2\Omega U\right)$ and ${\vec e}^{\vec\Omega}_{/\!/}$ is the unit vector along the rotation axis (${\vec e}^{\Omega}_{\perp}$ is respectively the perpendicular one). We have introduced the linear and non-linear Rossby numbers and the Reynolds { number}
\begin{equation}
R_{\rm o}=\frac{1}{2\Omega T}\equiv\frac{\sigma}{2\Omega},\,R_{\rm o}^{\rm NL}=\frac{U}{2\Omega\,L}\quad\hbox{and}\quad R_{\rm e}=\frac{UL}{\nu}.
\end{equation}
Then, neglecting the buoyancy force, different regimes can be identified.
\begin{itemize}
\item When $R_{\rm o}\!<\!\!<\!1$, $R_{\rm o}^{\rm NL}\!<\!\!<\!1$ and $R_{\rm e}\!>\!\!>\!1$, we identify the geostrophic equilibrium between the Coriolis acceleration and the pressure gradient.
\item When $R_{\rm o}\le1$ and $R_{\rm o}^{\rm NL}\!<\!\!<\!1$, we get linear progressive inertial waves, superimposed to the geostrophic equilibrium state. 
\item When $\left\{R_{\rm o},R_{\rm o}^{\rm NL}\right\}\le1$ and $R_{\rm e}\!>\!\!>\!1$,  we get the so-called "wave turbulence" regime, where the rotating strongly anisotropic turbulent flow can be described as a system of inertial waves in interaction. In this regime, thanks to non-linearities, transfers of energy occur between the various scales of the flow.
\item When $\left\{R_{\rm o},R_{\rm o}^{\rm NL}\right\}>1$ and $R_{\rm e}\!>\!\!>\!1$, we recover the case of the "classical" turbulence perturbed by rotation, with a characteristic time-scale that is larger than or of the order of the turbulent convective turnover time $L/U$. The turbulent flows are close to the assumption of homogeneous and isotropic turbulence if $R_{\rm o}^{\rm NL}\!>\!\!>\!1$. Finally, turbulence is coupled to super-inertial GIWs ($R_{\rm o}>1$) that are evanescent in convection zones.  
\end{itemize}

In stellar convection zones, characteristic dimensionless numbers can be qualitatively evaluated using the scaling-law for the convective velocities proposed by Brun (2014) 
\begin{equation}
u_{c}\approx\left[\frac{L_{*}}{{\overline \rho}_{\rm CZ}R^{2}}\right]^{1/3},
\label{Vc}
\end{equation}
where $L_{*}$ is the luminosity of the star, $R$ its radius and ${\overline \rho}_{\rm CZ}$ the mean density of the studied convective region. Then, the convective Rossby number can be evaluated as
\begin{equation}
R_{\rm o}^{NL;c}\equiv\frac{{\mathcal W}_{c}}{2\Omega}\approx\frac{u_{c}}{2\Omega D_{c}},
\end{equation}
where ${\mathcal W}_{c}$ is the convective vorticity and $D_{c}$ the thickness of the convective zone.}

\subsubsection{The asymptotic case of slow rotation} 

{ We consider the regime with $R_{\rm o}^{\rm NL}\!>\!\!>\!1$ discussed above, i.e. the case where rotation only weakly affects turbulence. This regime has been previously studied in the literature \citep[][]{SamadiGoupil2001,Samadi2003b,Samadietal2003a,BelkacemMathisetal2009,Samadietal2010}. We follow the method of \cite{SamadiGoupil2001} and \cite{Samadi2003b,Samadietal2003a,Samadietal2010}, and we introduce} the volumetric excitation rate, which is obtained from an integration by part of Eq.~(\ref{FA1}):
\begin{equation}
{\mathcal P}_{\rm R}=\frac{\pi^3}{I}\int_{0}^{M}{\rm d}m\left\{\,F_{\rm kin}\, L_{c}^{4}\int_{0}^{\infty}\,
\frac{{\rm d}k}{k^2}\left[\frac{16}{15}
\left(\frac{{\rm d}^{2}W}{{\rm d}z^{2}}\right)^{2}{\tilde{\mathcal S}}\left(k,\sigma\right)\right]\right\},
\label{Pf}
\end{equation}
where
\begin{equation}
{\tilde{\mathcal S}}\left(k,\sigma\right)=w_{\rm c}^{-3}L_{\rm c}^{-4}\left[E\left(k\right)\right]^{2}G_{k}\left(\sigma\right).
\label{Pf-in}
\end{equation}
{ In these equations, $w_{\rm c}$ is the rms value of the vertical component of the convective velocity ${\vec u}_{\rm c}$, $F_{\rm kin}=\frac{1}{2}{\overline\rho}w_{\rm c}^{3}$ the vertical flux of kinetic energy, $L_{\rm c}$ the characteristic length of convective flows, ${\rm d}m$ the infinitesimal mass element and $I$ the mass-mode.} Moreover, $E\left(k\right)$ is the kinetic energy spectrum associated with the turbulent convective velocity field, where $k$ is the eddy wavenumber { in the Fourier space}, and $G_{k}\left(\sigma\right)$ describes the time-dependent part of the turbulent spectrum, which models the correlation time-scale of an eddy with a wavenumber $k$. { As pointed by \cite{Samadi2011}, { the separation between turbulent velocities and waves and} this expansion of the turbulent source function can only be applied to isotropic turbulence { in the regime of slow rotation}, and we will see in Sec. 3.2.3 that in the case of rapid rotation { they} must be abandoned.} For a quantitative estimate of the amplitude, $E$ and $G_{k}$ can be computed using 3D numerical simulations \cite[e.g.][]{Samadi2003b,Belkacemetal2009,Rogersetal2013}. 
 
Until Sect. 3.2, the action of rotation has been taken into account for the modification of waves' velocity field only. { In particular, we have demonstrated that super-inertial GIWs are less and less evanescent when their Rossby number ($R_{\rm o}$) is decreased. This increases the possibility of couplings between GIWs and turbulent eddies. In fact, the weaker $p\left(R_{\rm o}\right)$ for a wave, the stronger the possibility for its coupling with an eddy of characteristic length-scale $1/k_{\perp}$.} In Eq. (\ref{Pf}), we  introduced $w_c$, the rms value of the vertical component of the convective velocity (Eq. \ref{Vc}). However, { while it provides us the strength of the vertical flux of kinetic energy transported by convective flows, it does not take their modification by rotation }into account. { We thus have to consider the evolution of the kinetic energy spatial spectrum ($E\left(k\right)$) as a function of the non-linear Rossby number ($R_{\rm o}^{\rm NL}$). Indeed, while in the non-rotating case, we can approximate the spectrum using a Kolmogorov-like form $E\left(k\right)\propto k^{-5/3}$, it becomes steeper as soon as rotation is increased: $E\left(k\right)\propto k^{-\beta\left(R_{\rm o}^{\rm NL}\right)}$ \citep[see e.g.][]{Zhou1995,Morizeetal2005}.}
{ Therefore, it is necessary to get spectra of rotating turbulent flows using numerical simulations and laboratory experiments to take into account the action of the Coriolis acceleration on convective flows. 

The rate of energy injection in GIWs is also a function of their time-correlation with turbulent eddies \citep{Samadietal2010}. On one hand, it is maximum for $P=2\pi/\sigma\ge T_{c}$, where $T_{c}$ is the characteristic eddy turn-over time and $P$ is the period of the modes. On the other hand, it decreases strongly when $P<T_{c}$. Since rotation affects general properties of turbulent eddies, it will also affect the time-correlation function.}

{ To evaluate how turbulent convective flows are influenced by rotation, we can use results obtained in numerical simulations and laboratory experiments. We introduce the Rayleigh number, $R_{\rm a}=\alpha_{T}g\Delta T D^3/\kappa\nu$, that compares the strength of buoyancy and diffusion in convective regions, and the Ekman number, $E=\nu/2\Omega D^2$, that compares the viscous force and the Coriolis acceleration. $\alpha_T$ is the fluid's thermal expansion coefficient, $g$ the gravity, $\Delta T$ the temperature difference across the convective region and $D$ its thickness. $\nu$ and $\kappa$ are the fluid viscous and thermal diffusivities respectively. \cite{Kingetal2012} and \cite{Julienetal2012} showed that the control parameter of the problem is a transition Rayleigh number that scales as a power-law of the Ekman number, i.e. $R_{\rm a;t}\sim E^{\gamma}$ (\cite{Julienetal2012} gives $\gamma=-8/5$ while \cite{Kingetal2012} predict $\gamma=-3/2$). At a given rotation rate, if $R_{\rm a}>R_{\rm a; t}$ the convection is weakly modified by rotation while cases where $R_{\rm a}<R_{\rm a; t}$ are in the rapidly rotating regime. A complete discussion of the properties of the flows can be found in \cite{Kingetal2013} (see also \cite{Julienetal2012b}).}

{ The modification of the amplitude of super-inertial GIWs in the slowly rotating case thus results from the modification of their propagation in convection zones by the Coriolis acceleration (Sect. 2.3.) and of its impact on convective eddies.}

\subsubsection{ The asymptotic case of rapid rotation} 
{ In the case of rapid rotation, as explained in Sect. 3.2.1, we cannot separate turbulence from propagative inertial waves (which correspond to sub-inertial GIWs in the radiation zones). Indeed, turbulent flows, which become highly anisotropic, can be understood as non-linearly interacting inertial waves. Following \cite{Senetal2012} \citep[see also e.g.][]{SmithWaleffe1999,Galtier2003,Cambonetal2004,Bourouibaetal2012}, we introduce the wave-vector expansion
\begin{equation}
{\vec k}=k\,{\vec {\widehat e}}_{\vec k}={\vec k}_{/\!/}^{\vec \Omega}+{\vec k}_{\perp}^{\vec \Omega}=k_{/\!/}^{\Omega}\,{\vec {\hat e}}^{\Omega}_{/\!/}+k_{\perp}^{\Omega}\,{\vec {\hat e}}^{\Omega}_{\perp},
\end{equation}
which has been decomposed along the directions respectively parallel and orthogonal to the rotation axis. Note that for $\alpha=\pi/2$ (see Fig. \ref{Fig1MN}), we have
\begin{equation}
{\vec k}_{/\!/}^{\vec \Omega}\!=\!k_{/\!/}^{\Omega}\left(\cos\Theta\,{\vec {\hat e}}_{z}+\sin\Theta\,{\vec {\hat e}}_{y}\right)\,\,\hbox{and}\,\,{\vec k}_{\perp}^{\vec \Omega}\!=\!k_{\perp}^{\Omega}\left(\sin\Theta\,{\vec {\hat e}}_{z}-\cos\Theta\,{\vec {\hat e}}_{y}\right),
\end{equation}
which correspond to $k_{V}+k_{\perp}{\tilde\delta}\left(R_{\rm o}\right)=k_{/\!/}^{\Omega}\cos\Theta+k_{\perp}^{\Omega}\sin\Theta$ and $k_{\perp}=k_{/\!/}^{\Omega}\sin\Theta-k_{\perp}^{\Omega}\cos\Theta$ for inertial waves described in Sect. 2.3.2. Then, the general solution for the velocity field ($\vec u$) of propagative inertial waves in the convection zone can be expanded \citep[see][]{SmithWaleffe1999}:
\begin{equation}
{\vec u}\left(\vec x,t\right)=\sum_{\vec k,s}b_{s}\left(\vec k\right){\vec h}_{s}\left(\vec k\right)\exp\left[i\left(\vec k\cdot\vec x-\sigma_{s}\,t\right)\right],
\label{v_rapid}
\end{equation}
where
\begin{equation}
{\vec h}_{s}={\vec {\hat e}}_{k}\times\left({\vec {\hat e}}_{k}\times{\vec {\hat e}}^{\Omega}_{/\!/}\right)+i\,s\,\left({\vec {\hat e}}_{k}\times{\vec {\hat e}}^{\Omega}_{/\!/}\right)
\end{equation}
is the Craya-Herring helical basis and
\begin{equation}
\sigma_{s}=s2{\vec\Omega}\cdot{\vec {\hat e}}_{k}=s2{\Omega}\frac{k_{/\!/}^{\Omega}}{\sqrt{\left(k_{/\!/}^{\Omega}\right)^2+\left(k_{\perp}^{\Omega}\right)^2}}\quad\hbox{with}\quad s=\pm 1.
\end{equation}
Waves are circularly polarised and they propagate with velocities that are normal to $\vec k$ and rotate around it during their propagation with a group velocity ${\vec v}_{\rm g}=s{\vec k}\times\left(2{\vec\Omega}\times{\vec k}\right)/\vert{\vec k}\vert^{3}$. Waves with a negative helicity ($s=1$) propagate upward (${\vec v}_{\rm g}\cdot\vec\Omega>0$), while those with a positive helicity ($s=-1$) travel downward (${\vec v}_{\rm g}\cdot\vec\Omega<0$).

If we adopt the framework of the weak inertial-wave turbulence theory \citep[e.g.][]{Galtier2003}, we substitute Eq. (\ref{v_rapid}) in Eq. (\ref{Eq_NL}) which leads to the following equation for $b_{s}\left(\vec k\right)\equiv b_{s_k}$:
\begin{eqnarray}
\left(\partial_{t}+\nu\, {\vec k}^{2} \right)b_{s_k}=R_{\rm o}^{\rm NL}\sum_{{\vec k}+{\vec p}+{\vec q}=0}^{\sigma_{s_k}+\sigma_{s_p}+\sigma_{s_q}=0}\sum_{s_p,s_q}\left\{C_{{\vec k},{\vec p},{\vec q}}^{{s_k},{s_p},{s_q}}b_{s_p}^{*}b_{s_q}^{*}\right.\nonumber\\
{\left.\times\exp\left[i\left(\sigma_{s_k}+\sigma_{s_p}+\sigma_{s_q}\right)t\right] \right\}},
\end{eqnarray}
with
\begin{equation}
C_{{\vec k},{\vec p},{\vec q}}^{s_k,s_p,s_q}=\left(s_q\,q-s_p\,p\right)\left[\left({\vec h}_{s_k}^{*}\times{\vec h}_{s_p}^{*}\right)\cdot{\vec h}_{s_q}^{*}\right],
\end{equation}
where $^{*}$ indicates a complex conjugate. The non-linear terms on the right-hand side correspond to triadic interactions and are of the order of $R_{\rm o}^{\rm NL}$. In the rapidly rotating case where $R_{\rm o}^{\rm NL}$ is small, the inertial waves oscillate on a rapid time-scale $\left(2{\Omega}\right)^{-1}$ while their amplitudes ($b_{s_k}$) evolve on the slow time-scale $\left({R_{\rm o}^{\rm NL}\Omega}\right)^{-1}$. In this framework, the terms $\exp\left[i\left(\sigma_{s_k}+\sigma_{s_p}+\sigma_{s_q}\right)t\right] $ with $\sigma_{s_k}+\sigma_{s_p}+\sigma_{s_q}\ne0$ are rapidly oscillating but they average out to zero when one considers the long timescale of order $(R_{\rm o}^{\rm NL}\Omega)^{-1}$. This leads us to only consider near resonance configurations in which
\begin{equation}
{\vec k}+{\vec p}+{\vec q}=0\quad\hbox{and}\quad \sigma_{s_k}+\sigma_{s_p}+\sigma_{s_q}={\mathcal O}\left(R_{\rm o}^{\rm NL}\right).
\end{equation}
If we assume the Waleffe's instability assumption \citep{SmithWaleffe1999}, these equations allow us to understand the mechanism of transfer of energy towards 2D vortical modes (with ${\vec k}^{\Omega}_{/\!/}=0$) that is responsible for the formation of strong columnar flows observed in rapidly rotating fluids in laboratory experiments and in numerical simulations.   

However, as pointed by \cite{Senetal2012}, it is necessary to go beyond the above formalism. Indeed, in realistic flows there are energy transfers from 3D to 2D modes but also from 2D to 3D modes that cannot be described by the weak inertial-wave turbulence theory. Following \cite{Senetal2012}, we decompose the velocity field in the convective region as a superposition of 3D (or wave modes) 
and of slow 2D vortical modes
:
\begin{equation}
{\vec u}\left({\vec k}\right)\!=\!
\begin{cases}
{\vec u}_{\rm 3D}\left({\vec k}\right)\hbox{ with $\vec k\in W_{k}$}\\
{\vec u}_{\rm 2D}\left({\vec k}_{\perp}^{\Omega}\right)={\vec u}_{\perp}\left({\vec k}_{\perp}^{\Omega}\right)+w\left({\vec k}_{\perp}^{\Omega}\right){\vec {\hat e}}_{/\!/}^{\Omega}\hbox{ with $\vec k\in V_{k}$}
\end{cases}
\end{equation}
where
\begin{equation}
\begin{cases}
W_{k}\equiv\left\{\vec k\hbox{ such that $\vert{\vec k}\vert\ne0$ and $k_{/\!/}^{\Omega}\ne0$}\right\}\\
V_{k}\equiv\left\{\vec k\hbox{ such that $\vert{\vec k}\vert\ne0$ and $k_{/\!/}^{\Omega}=0$}\right\}
\end{cases}.
\end{equation}
The total kinetic energy is expanded as
\begin{equation}
E=\sum_{\vec k}\vert{\vec u}\left(\vec k\right)\vert^2/2=E_{\rm 3D}+E_{\rm 2D}=E_{\rm 3D}+\left(E_{\perp}+E_{w}\right)
\end{equation}
where
\begin{equation}
\begin{cases}
E_{\rm 3D}=\sum_{\vec k\in W_{\vec k}}\left|{\vec u}_{\rm 3D}\left({\vec k}\right)\right|^{2}/2\\
E_{\rm 2D}=\sum_{{\vec k}_{\perp}^{\Omega}}\left|{\vec u}_{\rm 2D}\left({\vec k}_{\perp}^{\Omega}\right)\right|^{2}/2\\
E_{\perp}=\sum_{{\vec k}_{\perp}^{\Omega}}\left|{\vec u}_{\perp}\left({\vec k}_{\perp}^{\Omega}\right)\right|^{2}/2\\
E_{w}=\sum_{{\vec k}_{\perp}^{\Omega}}\left| w\left({\vec k}_{\perp}^{\Omega}\right)\right|^{2}/2.
\end{cases}
\end{equation}
By multiplying the spectral form of the momentum equation by ${\vec u}^{*}\left(\vec k\right)$, we can derive the two coupled differential equations for the total energy in 3D inertial modes and in 2D slow modes \citep[see][]{Senetal2012}:
\begin{eqnarray}
\frac{{\rm d}}{{\rm d}t}E_{\rm 3D}&=&\Pi_{\rm 2D\rightarrow3D}-\Pi_{\rm 3D}+{\varepsilon}_{\rm 3D}\nonumber\\
\frac{{\rm d}}{{\rm d}t}E_{\rm 2D}&=&-\Pi_{\rm 2D\rightarrow3D}-\Pi_{\rm 2D}+{\varepsilon}_{\rm 2D}.
\end{eqnarray}
In these equations, $\varepsilon_{j}$, where $j=\left\{{\rm 3D},{\rm 2D}\right\}$, corresponds to the forced energy injection into $j$-modes and $\Pi_{j}$ is the energy that is transferred to small scales and dissipated per unit of time,  balancing $\varepsilon_{j}$.
Finally, $\Pi_{\rm 2D\rightarrow3D}$, which is of the order of $R_{\rm o}^{\rm NL}$, is the flux of energy that is transferred from 2D to 3D modes when it is positive (and from 3D to 2D modes when it is negative). 

When applied to numerical simulations of rapidly rotating turbulent flows \citep[e.g.][]{Senetal2012}, these equations allow us to isolate the energy contained in 3D and in 2D modes and thus their respective strength. Moreover, in \cite{Mininnietal2011} and \cite{Senetal2012}, the authors tested various types of forcing and demonstrated that, in most of their simulations, $\Pi_{\rm 2D\rightarrow3D}$ is negative for small values of $k_{/\!/}^{\Omega}$, which indicates that the energy goes from 3D modes towards 2D ones for larger scales, and positive for large values of $k_{/\!/}^{\Omega}$ \citep[energy is transferred to 3D modes for smaller scales; see also Fig. 12 in][]{Senetal2012}. Moreover, to identify wave vectors in rapidly rotating turbulent regions, it is interesting to define the axisymmetric energy spectrum
\begin{equation}
e\left(k_{\perp}^{\Omega},k_{/\!/}^{\Omega}\right)=e(k,\theta_{k})=\frac{1}{2}\sum_{\begin{subarray}{l} {k_{\perp}^{\Omega}<\vert {\vec k}\times{\vec {\hat e}}_{/\!/}^{\Omega} \vert < k_{\perp}^{\Omega}+1} \\ {k_{/\!/}^{\Omega}\le \vert{\vec k}\cdot {\vec {\hat e}}_{/\!/}^{\Omega}\vert < k_{/\!/}^{\Omega}+1} \end{subarray}}\vert {\vec u}\left(\vec k\right)\vert^2,
\end{equation}
where $\theta_{k}$ is the colatitude with respect to the rotation axis in the Fourier space. We identify
\begin{equation} 
E_{\rm 2D}=\sum_{k_{\perp}^{\Omega}}e\left(k_{\perp}^{\Omega},k_{/\!/}^{\Omega}=0\right)=\sum_{k}e\left(k,\theta_{k}=\pi/2\right)
\end{equation}
and
\begin{equation}  
E_{\rm 3D}=\sum_{k}\left\{E\left(k\right)-e\left(k,\theta_{k}=\pi/2\right)\right\}.
\label{3D}
\end{equation}
An interesting application of this diagnosis can be found in \cite{Leonietal2013} in which the strength of inertial waves in a rotating turbulent flow has been examined. Moreover, in \cite{SJP2013}, a weak-wave turbulence theory for rotationally constrained slow inertial waves is built. In the case of non-helical dynamics, it leads to a spectrum of the form 
\begin{equation}
e_{\vec k}=e\left(k_{\perp}^{\Omega},k_{/\!/}^{\Omega}\right)=\sum_{s_{k}}b_{s_{k}}b_{s_{k}}^{*}\propto{k_{\perp}^{\Omega}}^{-3}{k_{/\!/}^{\Omega}}^{-1},
\end{equation}
which has been observed in several numerical simulations \citep[e.g.][]{TM2012,MRP2012}. In the helical case, the scaling $\propto{k_{\perp}^{\Omega}}^{-3}$ is { observed} \citep[see also][]{MP2010}. { This anisotropic kinetic energy spectrum is directly related to the amplitude coefficients of inertial waves ($b_{s}\left(\vec k\right)$; see Eqs. \ref{v_rapid} and \ref{3D}) that become sub-inertial GIWs in the adjacent radiative zone. It corresponds to Eqs. (\ref{Pf}) and (\ref{Pf-in}) derived in the slowly rotating case, the important differences being that in the case of rapid rotation waves and turbulent flows cannot be separated as in the slowly rotating regime and that turbulence becomes anisotropic.}

In the context of the stochastic excitation of GIWs in rapidly rotating stars, these diagnoses are very important. First, as it is shown in Figs. \ref{Fig3MN} and \ref{Fig5MN}, 3D inertial modes that propagate in turbulent convection zones, convert into sub-inertial GIWs in the adjacent radiative regions. { Next, the turbulent structures associated to 2D modes, in which the energy propagates along the rotation axis, will also excite GIWs because of their penetration at the interfaces between convection and radiation \citep[e.g.][]{TakehiroLister2001}}. In this context, it is mandatory to go beyond the traditional approximation, the non-traditional terms being those that allow us to properly couple the gravity and inertial modes (see Fig. \ref{Fig3MN}).\\

In the two above subsections, we have discussed the way in which rotation deeply impacts the nature of turbulent flows that excite stochastically GIWs in stellar interiors. On one hand, in the case of slow rotation, turbulence is only weakly perturbed by rotation and is coupled to super-inertial GIWs, which are evanescent in convective zones. When rotation is increased, they are less and less evanescent until $R_{\rm o}=1$. On the other hand, in the case of rapidly rotating stars, turbulent flows are deeply modified by rotation and they become intrisically coupled with inertial waves, which corresponds to sub-inertial GIWs. This theoretical picture shows the importance of computing high-resolution numerical simulations of rapidly rotating turbulent convective stellar regions with adjacent radiative zones in a near future to test our scenario as a function of $R_{\rm o}$ and of $R_{\rm o}^{\rm NL}$ for various stellar types \citep[e.g.][]{Browningetal2004,Ballotetal2007,Brownetal2008,Mattetal2011,Brunetal2011,Rogersetal2013,ABM2013}

\begin{figure}[t!]
\centering  
\includegraphics[width=0.5\textwidth]{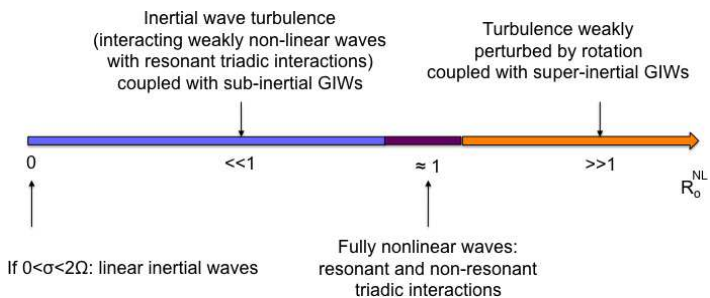}      
\caption{Wave-turbulence couplings as a function of the non-linear Rossby number $\left(R_{\rm o}^{\rm NL}\right)$. Resonant excitation is obtained for $R_{\rm o}^{\rm NL}\approx R_{\rm o}$.}
\label{Fig6MN}
\end{figure}

}

\section{Discussion and conclusions}\label{discuss}

In this work, we have formally demonstrated that rotation, through the Coriolis acceleration, modifies the stochastic excitation of gravity waves and GIWs, the control parameters being the wave's Rossby number $R_{\rm o}=\sigma/2\Omega$ { and the non-linear Rossby number $R_{\rm o}^{\rm NL}$ of convective turbulent flows}. On one hand, in the super-inertial regime ($\sigma>2\Omega,\hbox{ {\it i.e.} }R_{\rm o}>1$), the evanescent behaviour of GIWs above (below) the external (internal) turning point becomes increasingly weaker as the rotation speed grows until $R_{\rm o}=1$. { Simultaneously, the turbulent energy cascade towards small scales is slowed down.} The coupling between super-inertial GIWs and given turbulent convective flows is then amplified as described in Eq. (\ref{FA1}). On the other hand, in the sub-inertial regime ($\sigma<2\Omega,\hbox{ {\it i.e.} }R_{\rm o}<1$), GIWs become propagative inertial waves in the convection zone. { In the case of rapid rotation, turbulent flows, which become strongly anisotropic, result from their non-linear interactions. Sub-inertial GIWs that correspond to propagating inertial waves in convection zones are then intrinsically and strongly coupled to rapidly rotating turbulence as discussed in Sect. 3.2.3. These different regimes are summarised in Fig. \ref{Fig6MN}.}

Such effects are of great interest for asteroseismic studies of rotating stars since gravity wave and GIW amplitudes are thus expected { to} be stronger in rapidly rotating stars, { a conclusion that is supported by recent numerical simulations \citep{Rogersetal2013}}. For example, until recently, stochastically excited gravity waves were thought to be of too low amplitude to be detected even with space missions such as CoRoT \citep{Samadietal2010}. The discovery of stochastically excited GIWs in the rapid rotator HD\,51452 \citep{Neineretal2012} proved that such waves can be detected. Our results show how the amplitude of these waves can be enhanced thanks to rotation. The interpretation of observed pulsational frequencies in rapid rotators should take this into account. In particular,  oscillations observed in $\beta$\,Cep and Slowly Pulsating B (SPB) stars should not be systematically attributed to the $\kappa$-mechanism, as it was done until now, if the star rotates fast. For example, the GIWs observed in HD\,43317 \citep{Papicsetal2012} might be of stochastic origin and might have been enhanced by rapid rotation. This is of course especially true for Be and Bn stars.

Moreover, the related transport of angular momentum, which until now was believed to become less efficient because of GIWs equatorial trapping in the sub-inertial regime \citep{Mathisetal2008,Mathis2009,MdB2012}, may be sustained thanks to the stronger stochastic excitation by turbulent convective flows. This may have important consequences for example for rapidly rotating young low-mass stars \citep{CharbonnelTalon2008,Charbonneletal2013} and active intermediate-mass and massive stars such as Be stars. For example, \cite{Neineretal2013} proposed that the outburst of the Be star HD\,49330 observed by CoRoT \citep{Huatetal2009} is due to the deposit of angular momentum by GIWs just below the surface \citep[see also][]{Lee2013}. 

Our prediction should now be confronted to realistic numerical simulations of  stochastic excitation of GIWs in stellar interiors \citep[e.g.][]{BMT2011,Rogersetal2012,Rogersetal2013,ABM2013}, to laboratory experiments \citep{Perrardetal2013}, as well as to a larger statistical sample of observed pulsating stars with detected GIWs. Moreover, a global formalism to treat the problem in the spheroidal geometry corresponding to rotating stars must be built in a near future.

\begin{acknowledgements}
{ We thank the referee for her/his remarks and suggestions that improved the original manuscript.} This work was supported by the French Programme National de Physique Stellaire (PNPS) of CNRS/INSU, the CNES-SOHO/GOLF grant and asteroseismology support at CEA-Saclay, the CNES-CoRoT grant at LESIA, and by the Programme Physique Th\'eorique et ses interfaces of CNRS/INP. Authors are grateful to A.-S. Brun, L. Alvan, R. A. Garcia, T. Rogers, M. Lebars, and D. Lecoanet for fruitful discussions and to P. Eggenberger, T. Decressin and M. Briquet for providing stellar models. 
\end{acknowledgements}

\bibliographystyle{aa}  
\bibliography{MNT2014} 

\end{document}